\documentclass[10pt]{iopart}
\usepackage{epsfig}
\usepackage{graphicx}
\begin{document}\hbadness=10000
\title{A comparison of statistical hadronization models}
\author{Giorgio Torrieri and Johann Rafelski}
\address{
Department of Physics, University of Arizona, Tucson, AZ 85721}
\date{December, 2002}
\begin{abstract}
\noindent
We investigate the sensitivity of fits of hadron spectra produced in heavy ion collisions to the choice
of statistical hadronization model.
We start by giving an overview of statistical model ambiguities,
and what they tell us about freeze-out dynamics.
We then use Montecarlo generated data to determine sensitivity to
model choice.
We fit the statistical hadronization models under consideration to RHIC data, and find that
a comparison $\chi^2$ fits can shed light on some presently
contentious questions.
\end{abstract}
\vspace{+0.5cm}
\section{Statistical hadronization models}
There is increasing interest to apply the statistical model  \cite{Fer50,Pom51,Hag65} in a description of both
particle yields  \cite{rafelski2002,PBM99,PBM01,PBMRHIC} and spectra \cite{burward-hoy,van-leeuwen,NA57,castillo,florkowski,ourspspaper,bugaev_freeze} produced in heavy ion collisions
in terms of a statistically hadronizing transversely expanding fireball.
The fitted parameters, and in particular the temperature, however,
have varied considerably,
ranging from as low as 110 MeV \cite{burward-hoy,van-leeuwen,NA57,castillo,bugaev_freeze}
to 140 MeV \cite{rafelski2002,ourspspaper}
to as high as 160 and 170 MeV
\cite{PBM99,PBM01,PBMRHIC,florkowski}.
This report examines these differences in terms of  the different possible approaches to statistical hadronization, and studies data sensitivity to model choice.

The most commonly used prescription for statistical emission of hadrons is the truncated
Cooper-Frye formula \cite{cooperfrye}.
\begin{equation}
\label{cooperfrye}
\left(E \frac{d N}{d^3 p}\right)_{cf} = \int p^{\mu} d^3 \Sigma_{\mu} f (p^{\mu} u_{\mu}, T, \lambda) \theta(p^{\mu} d^3 \Sigma_{\mu})+\left(E \frac{d N}{d^3 p}\right)_{Res}
\end{equation}
Where $p^{\mu}$ is the particle's four-momentum,
$u^{\mu}$ is the system's transverse flow,
T is the temperature,
$\lambda$ is the normalization and/or chemical potential, $f(E,T,\lambda)$ is the statistical distribution of the emitted particles in terms of
energy and conserved quantum numbers,
$\Sigma^{\mu}$ describes the hadronization geometry 
and $\theta(p^{\mu} d^3 \Sigma_{\mu})$ is a step function, which
corrects for inward emission  \cite{cf_trunc1,cf_trunc2}.
\begin{figure*}[tb]
\begin{center}
\centerline{\resizebox*{!}{0.247\textheight}{
\includegraphics{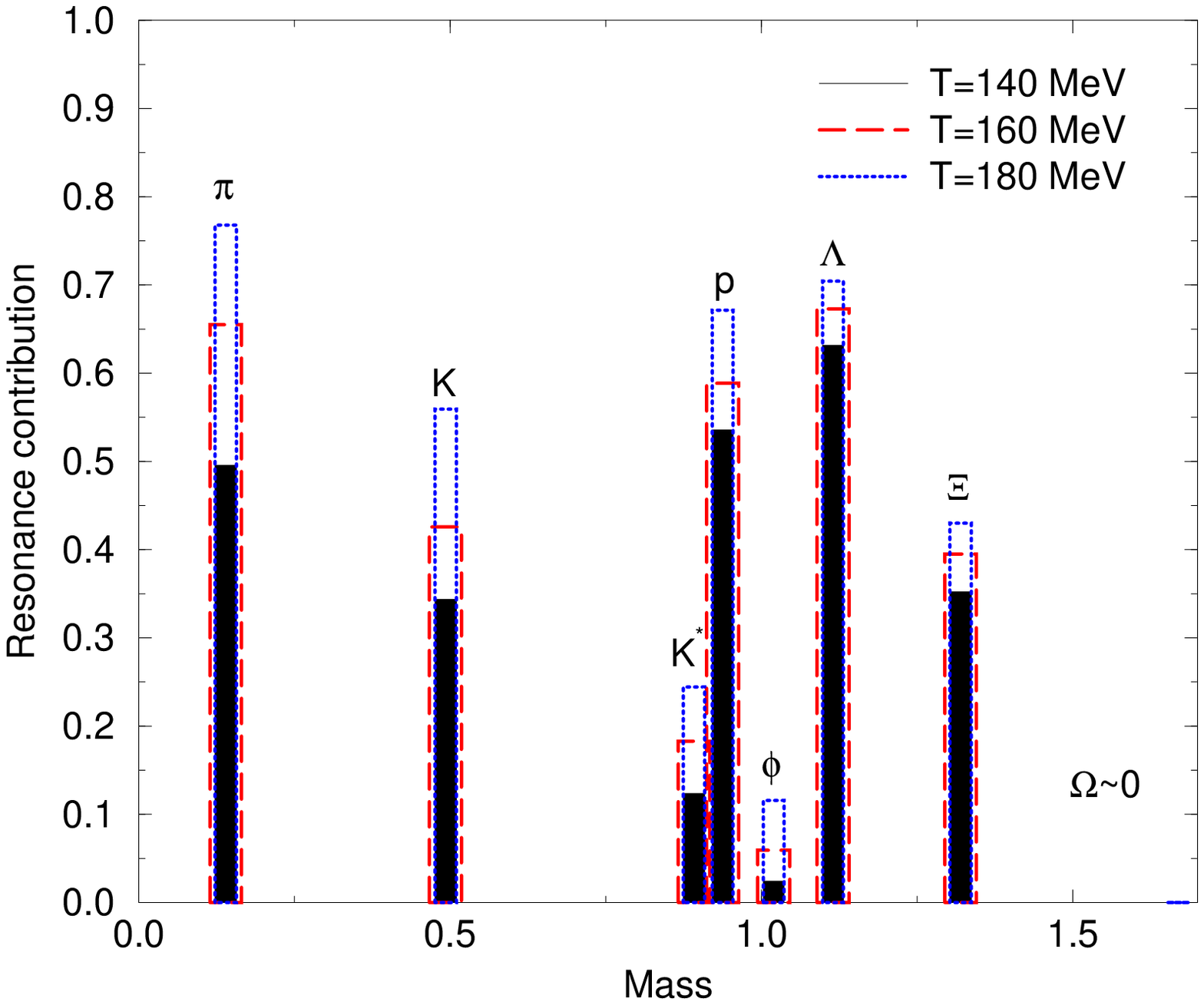}
}
\resizebox*{!}{0.247\textheight}{
\includegraphics{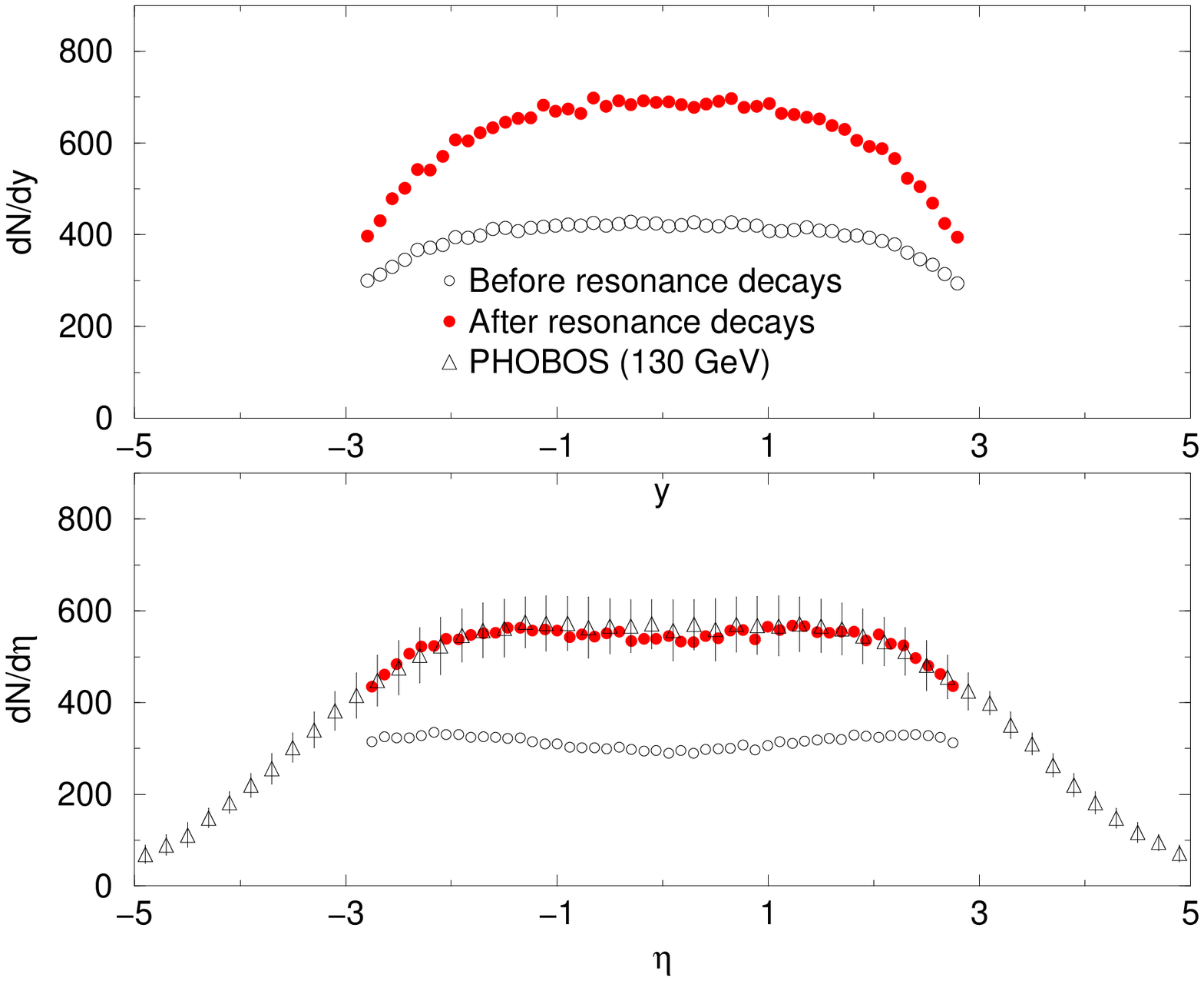}
}}
\end{center}
\caption{Left: Fraction of particles produced by resonances as a function
of Temperature, 
Right:  $dN_{tot}/d(y,\eta)$ arising from freeze-out of a Bjorken fluid, before (open circles) and after (solid circles) resonance decays.   The normalization is arbitrary, chosen to coincide with PHOBOS \cite{phobos} data (triangles)
\label{globals} }
\end{figure*}
The Cooper-Frye formula is appealing because it's the most general covariant
way of describing statistical particle emission.
However, its form is compatible with a variety of freeze-out scenarios.
\subsection{Resonances and normalization}
As Fig. \ref{globals} shows, the resonance contribution $(E \frac{d N}{d^3 p})_{Res}$ is a non-negligible
and strongly temperature dependent admixture to the total multiplicity.
Its role within particle spectra, however, is a matter of controversy.
If an interacting hadron gas phase re-equilibrates resonance decay
products \cite{bleicher},
it might be more realistic to not include resonances when analysing spectra  \cite{burward-hoy,van-leeuwen,NA57,castillo}.
Direct detection of resonances
\cite{fachini,vanburen,markert,friese} through invariant mass reconstruction, however, suggests that at least in some cases their feed-down has to be taken into account.
These resonances  will have a  different distribution than those produced directly, and their inclusion can shift the temperature by as much as 20 MeV \cite{sqm2001}.

If an incoherent system is considered, one can factor out the decay matrix elements (the angular distribution of decay products and lifetime), and arrive
at the distribution of ``daughter'' particles ($N$ with momentum $p$) from ``parent particles''
($N_R$ with momentum $p_R$)
by simply integrating over the available phase space
\begin{equation}
\left(E \frac{dN}{d^3 p}\right)_{res}=\int \prod_{j=2}^{N} \frac{d^3
p_{j}}{E_{j}}
\delta(p_{R}^{\mu}-p^{\mu}-\sum_{2}^{N} p_{j}^{\mu}) \left(E_R \frac{dN_R}{d^3 p_R}\right)
\label{rfrom}
\end{equation}
In general, this expression can get very complicated, and Montecarlo
integration \cite{mambo} becomes necessary.   For most cases considered here, where
there is one feed-down and two or three body decays, eq. \ref{rfrom} can be
integrated semi-analitically \cite{florkowski,ourspspaper,fireballspectra}.

While including resonances in this way is computationally demanding \cite{fireballspectra}, it can be implemented with no extra degrees of freedom.
It is therefore straightforward to check the effect of resonance inclusion
on $\chi^2$.
A large increase in statistical significance would be a strong indication that
resonance decay products do in fact emerge from the system with little
rescattering.
Otherwise, rescattering has to be taken into account.

The handling of resonance decays is connected with the way
particle spectra are normalized.
Several ways to normalize spectra are possible, and
the fitted temperature could change significantly depending on the chosen method,
since temperature becomes correlated to a different set of parameters.

The simplest approach is to treat normalization
of each particle as a free parameter \cite{burward-hoy,van-leeuwen,NA57,castillo}.
If the post-hadronization interacting hadron gas phase is long enough to allow
inelastic interactions to alter the chemical composition of the fireball, this
may be the correct approach.
In this case, temperature will be correlated to flow but not to normalization.\\
The arbitrary normalization means there is no way to include feed-down from resonances  without introducing many more parameters (the resonance normalizations) into the fit.  (If rescattering is significant, however, these parameters would be physically justified).

Alternatively, a common normalization volume, together with the introduction of flavor chemical potentials, can be used
 \cite{florkowski,ourspspaper}.
This approach, justified by the success in handling short-lived resonances through thermal models \cite{PBMRHIC}, is consistent with a scenario in which post-hadronization dynamics does not change particle abundances or distributions significantly.  It does not require extra degrees of freedom for resonances to be included.

However, feed-down from resonances  means the fitted temperature will be strongly correlated with
chemical potentials and reaction volume, in addition to transverse flow:
Most short-lived resonances have the same quark content as the lighter daughter particles, making their yield relative to the lighter particle
dependent on temperature only.
Hence, temperature will influence both absolute normalization (correlating it with
volume, chemical potentials) and the slope (correlating it with flow).

If chemical potentials are used, the eventuality of chemical
non-equilibrium raises the possibility that, additionally to chemical potentials, the flavor phase space occupancy parameter
$\gamma$ needs to be used for either strange \cite{PBM01,florkowski} or light 
\cite{rafelski2002,observing} flavors.
If these parameters are used, they will correlate temperature with chemical
potential even
more strongly, since, just like temperature, $\gamma$ affects particles
and antiparticles abundance in the same way.

\subsection{Freeze-out geometry and flow profile}
Several choices of the freeze-out geometry $\Sigma^{\mu}$ are possible.
This choice is important both for getting a full understanding of how freeze-out happens  \cite{resratios}
 and because it correlates with the shape
of particle spectra, and hence on parameters such as transverse flow and
temperature.

Measured RHIC rapidity distributions \cite{phobos,brahms}
indicate that around midrapidity Bjorken boost-invariance \cite{bjorken} holds.
This, together with cylindrical symmetry (appropriate for the most central collisions) 
constrains $d^3 \Sigma^{\mu}$ to be of the form \cite{fireballspectra,resratios}
\begin{eqnarray}
\Sigma^{\mu} = [t_f \cosh(y_L), r \cos(\theta),r\sin(\theta),t_f \sinh(y_L)]\\
p^{\mu} d\Sigma_{\mu}=(m_T \cosh(y-y_L)-\frac{\partial t_f}{\partial r} p_T \cos(\theta-\phi))r dr dy
\end{eqnarray}
where $y,\phi$ are the particle's rapidity and momentum directions, while $y_L,\theta$ are the longitudinal
rapidity and emission angle of the freeze-out hypersurface.
$t_f$ is the freeze-out time in a longitudinally
co-moving frame.

It is clear that several freeze-out models can be produced with different choices of $\partial t_f/\partial r$, giving
quantitatively different results.   
It is also clear that the choice of $\partial t_f/\partial r$ might
influence fitted parameters such as temperature, transverse velocity and
normalization.
Recently published fits use a variety of freeze-out geometries:
\begin{itemize}
\item \cite{burward-hoy,van-leeuwen,NA57,castillo} assume $\partial t_f/\partial r=0$, i.e. emission is independent of radius.
Emission happens in the same lab momentum time.
\item \cite{ourspspaper} fits $\partial t_f/\partial r$ as a free parameter, with the result being $\sim 1$.\\
This indicates explosive freeze-out, driven by negative vacuum pressure \cite{sudden}
\item In \cite{florkowski} freeze-out happens at the same
proper time $\tau$ throughout the fireball, ($d^3 \Sigma^{\mu} \propto u^{\mu}$). Freeze-out might have this form (``Hubble'' freeze-out) if dynamical effects are negligible during hadronization \cite{resratios}.
\end{itemize}

An additional physical uncertainty, which is found to be correlated to
flow and $\partial t_f/\partial r$ in fits, is the flow profile.
Hydrodynamics requires that each part of the fireball
volume will in general have a different density and transverse expansion rate.
For this reason, the integral over $d^3 \Sigma$ will in general span a range
of flows, weighted by density function $\rho(r)$.
\begin{equation}
\label{flowprof}
E \frac{dN}{d^3 p} = \int r dr (E-p_T \frac{dt_f}{dr}) f(T,y_T (r),\lambda) \rho(r)
\end{equation}
Just like the form of $d^3 \Sigma$, the flow profile depends
on the hadronization conditions.
For instance, assuming that freeze-out happens at a critical energy density
yields  $y_T \propto r$ \cite{shuryak}, while
assuming freeze-out happening at approximately the same time in the lab frame leads to a power distribution for the transverse expansion velocity
$v_T \propto r^{\alpha}$   \cite{hydroheinz}\\
One can approximate the flow profile with one ``average'' flow and assume 
 \cite{NA57,ourspspaper}
\begin{equation}
\label{1flow}
E \frac{dN}{d^3 p} \propto (E-p_T \frac{dt_f}{dr}) f(T,<y_T>,\lambda)
\end{equation}
However, this approach might result in a systematic shift of fitted parameters
such as flow and  $\partial t_f/\partial r$
\begin{figure*}[tb]
\begin{center}
\centerline{\resizebox*{!}{0.247\textheight}{
\includegraphics{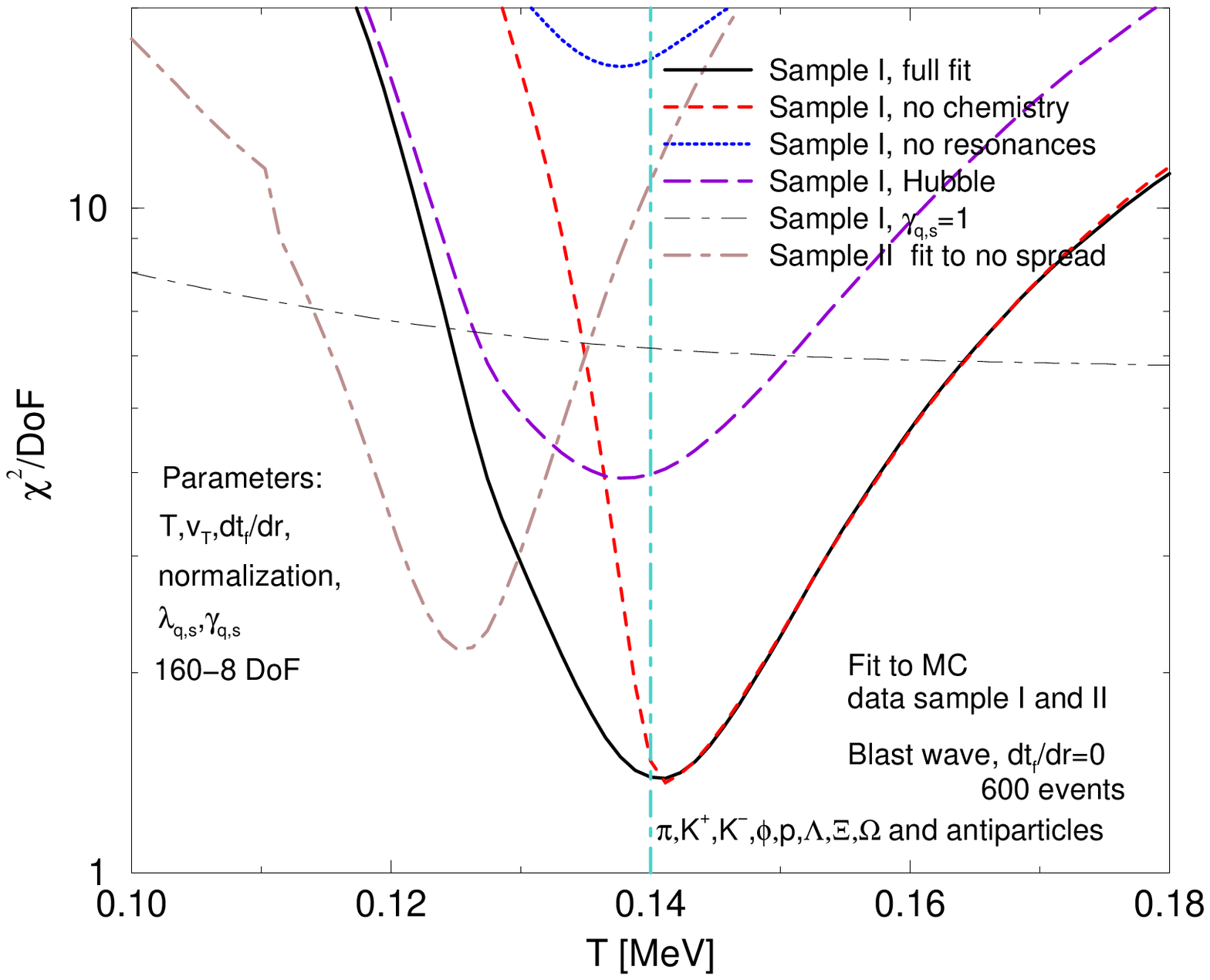}
}
\resizebox*{!}{0.247\textheight}{
\includegraphics{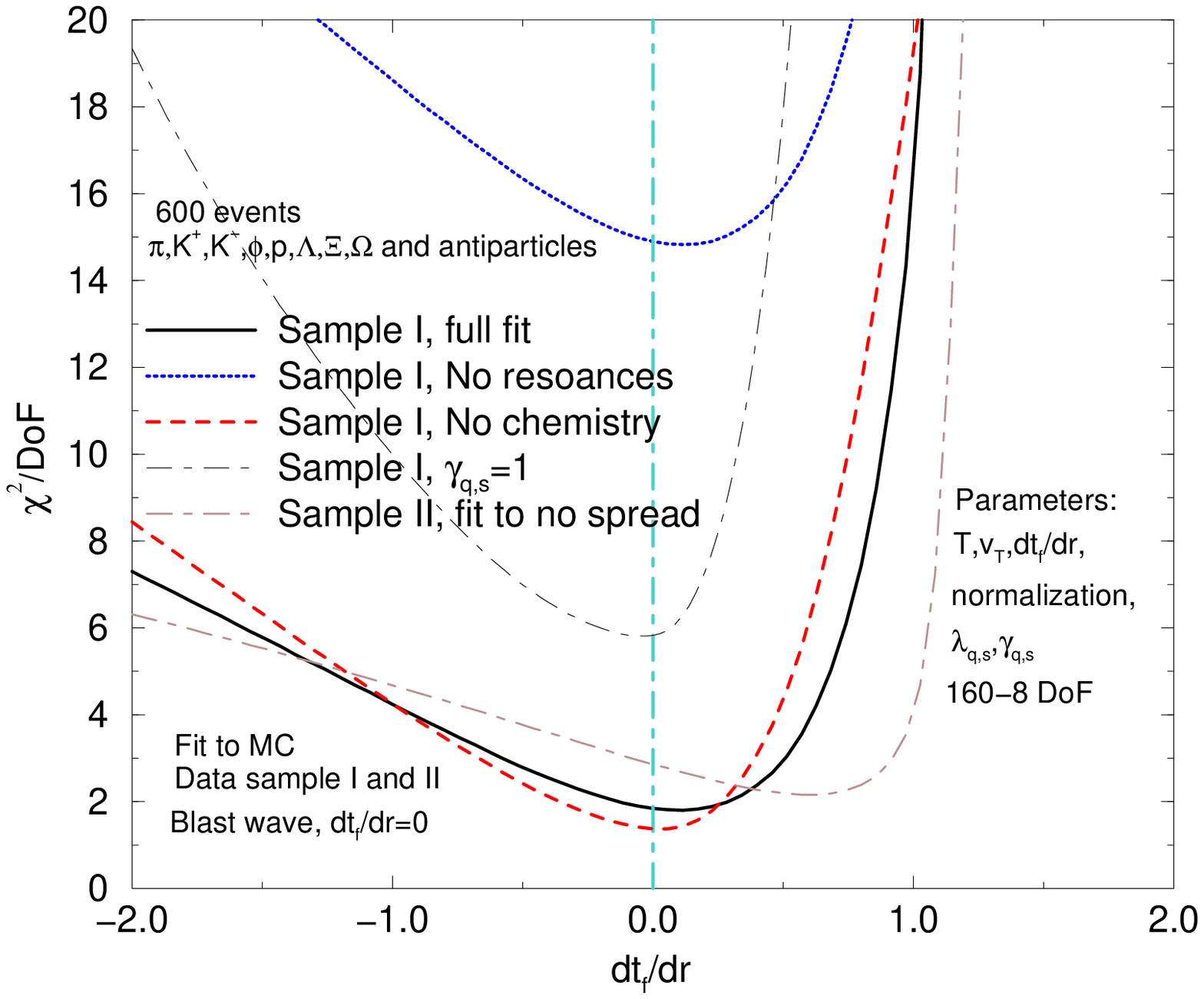}
}}
\end{center}
\caption{Results of fits to Montecarlo generated data samples I and II ($\frac{dt_f}{dr}=0$):\\
$\chi^2$ profiles for temperature (left) and fitted $\partial t_f/\partial r$ (right).   The models used in the fits are described in sections 1.1 and 1.2\\
Full fit includes fitted chemical potentials and $\gamma_{q,s}$, resonances and fitted $\partial t_f/\partial r$.\\
Last profile shows effect of fitting sample II using Eq.
\ref{1flow} (Sample I Full fit) \label{prof0} }
\end{figure*}
\section{Sensitivity to model choice}
The ambiguities presented in sections 1.1 and 1.2 mean that it is important
to study their effect on the statistical model's fitted parameters.
One way to do this is to use a Montecarlo to generate data according
to a particular freeze-out model, and to see what happens if the ``wrong'' model is used to perform the fit.
We have written a MonteCarlo program which can be used for this
purpose.
While a detailed description of the algorithm is outside the scope of this write-up
\cite{qm2002}, we shall provide a short summary of its features here.
An acceptance/rejection algorithm is used to generate particles
in a statistical distribution in the volume element's rest frame.
The accepted particles are then Lorentz transformed to the lab frame.
(Any flow and density profile, as well as any freeze-out surface can 
be accomodated).
Resonance decays are handled through Eq. \ref{rfrom}, using the MAMBO algorithm \cite{mambo} to generate points in phase space.
Output can be used to generate spectra or fed into a microscopic model such as uRQMD \cite{uRQMD}.

The Montecarlo output was used to produce the data points in Fig. \ref{globals}.
It can be seen that a boost-invariant statistical hadronization
can explain the global properties of the system such as $dN_{tot}/d\eta$.   It can also be verified that
the role of resonances is absolutely crucial.

We proceeded to generate three datasets of particles.
Each data set had a temperature of 140 MeV, a maximum transverse
flow of 0.55, and out of equilibrium chemistry ($\gamma_q=1.4,\gamma_s/\gamma_q=0.8$).
Generated particles include $\pi,K^{+},K^{-},p,\overline{p},\Lambda,\Xi,\Omega$ and their resonances.
The three samples differ in their choice of freeze-out geometry (specifically $\partial t_f/\partial r$) and flow profile:
\begin{description}
\item[Sample I]  $\partial t_f/\partial r=0$ and no flow profile, as fitted in \cite{NA57}
\item[Sample II] $\partial t_f/\partial r=0$ and a quadratic
flow profile, as fitted in \cite{burward-hoy,van-leeuwen,castillo}
\item[Sample III]  $\partial t_f/\partial r=1$, the boost-invariant analogue of \cite{ourspspaper}.
\end{description}
\begin{figure*}[tb]
\begin{center}
\centerline{\resizebox*{!}{0.247\textheight}{
\includegraphics{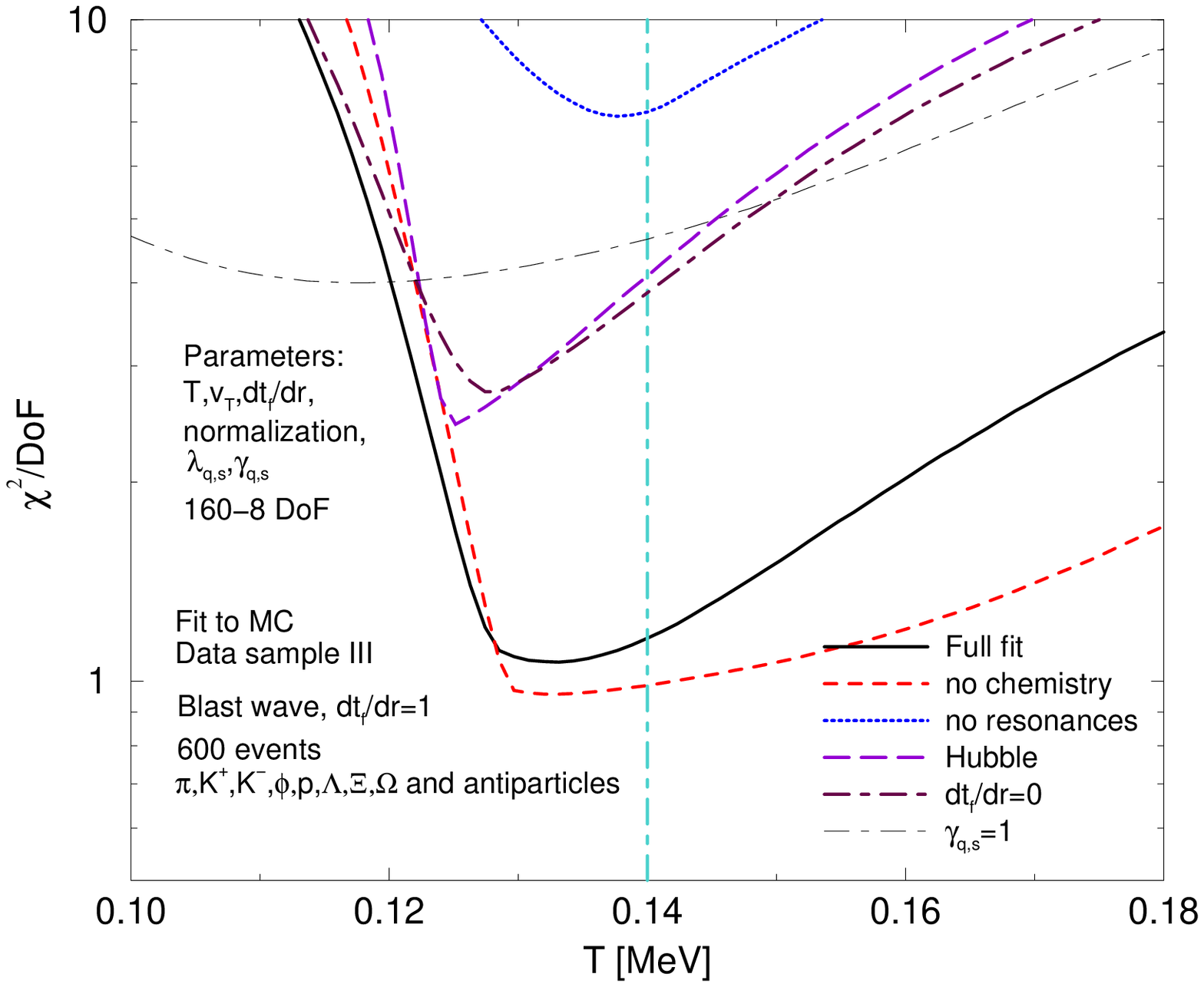}
}
\resizebox*{!}{0.247\textheight}{
\includegraphics{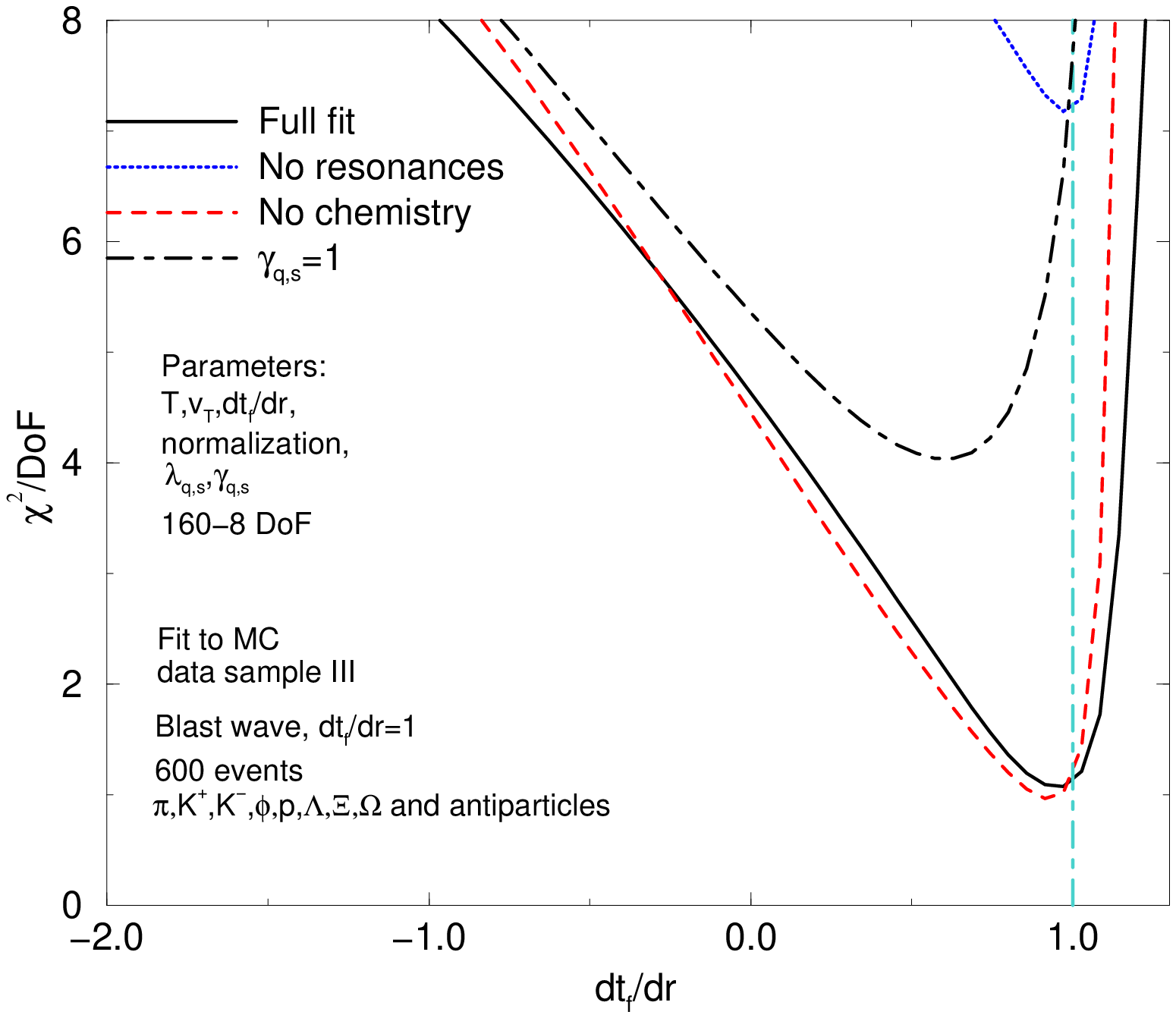}
}}
\end{center}
\caption{
Results of fits to Montecarlo generated data samples III ($\frac{dt_f}{dr}=1$):\\
$\chi^2$ profiles for temperature (left) and fitted $\partial t_f/\partial r$ (right).   The models used in the fits are described in sections 1.1 and 1.2\\
Full fit includes fitted chemical potentials and $\gamma_{q,s}$, resonances and fitted $\partial t_f/\partial r$.
\label{prof1} }
\end{figure*}
We have fitted the three samples to a variety of models, producing
$\chi^2/DoF$ profiles for freeze-out temperature and the $\partial t_f/\partial r$ parameter.
Fig. \ref{prof1} shows the profiles resulting in the fit to sample III while samples I and II are shown in Fig. \ref{prof0}.
A full chemistry model with resonances  (solid black line) seems to be equivalent (as far
as the position of the $\chi^2$ minimum and the value of $\chi^2/DoF$) 
to a fit in which normalization is particle-specific (dashed red line).
However, the chemical potential fit has greater statistical
significance since considerably more
degrees of freedom are required for arbitrary normalization.

If chemical potentials are included resonances
become essential since a fit with chemical potentials but no resonances
(dotted blue line) loses all statistical significance.
Similarly, the physical presence of non-equilibrium
($\gamma_{q,s} \neq 1$) means chemical potentials have to include
the non-equilibrium parameter for the fit to be meaningful (black dot-dashed line).
The freeze-out geometry does not seem to impact the temperature minimum that much.
However,  the correct freeze-out geometry can
be picked out by a comparison of fits to different models by choosing
the model with the lower $\chi^2/DoF$.
Moreover, data sensitivity to temperature is strongly affected by freeze-out
geometry:
Comparing the temperature profiles for different choices of $\partial t_f/\partial r$ (Figs. \ref{prof0} and \ref{prof1}) it is
apparent that the temperature $\chi^2$ minimum is more definite in
the $\partial t_f/\partial r=0$ case.
In the case of explosive freeze-out, the correlation
between temperature and other parameters in the fit (notably flow) increases, resulting in a shallow $\chi^2$ increase at larger than minimum temperatures.

Finally it should be noted that flow profile, freeze-out geometry and temperature appear to be
strongly correlated.
If data sample II is fitted with a distribution with no flow profile (such as Sample I) there is a non-negligible shift in both the fitted temperature and
$\partial t_f/\partial r$, and a small rise in $\chi^2/DoF$ (Fig. \ref{prof0}, brown dot-dashed line).
\section{Fit to RHIC data ($\sqrt{s_{NN}}=130 GeV$)}
\begin{figure*}[tb]
\begin{center}
\centerline{\resizebox*{!}{0.247\textheight}{
\includegraphics{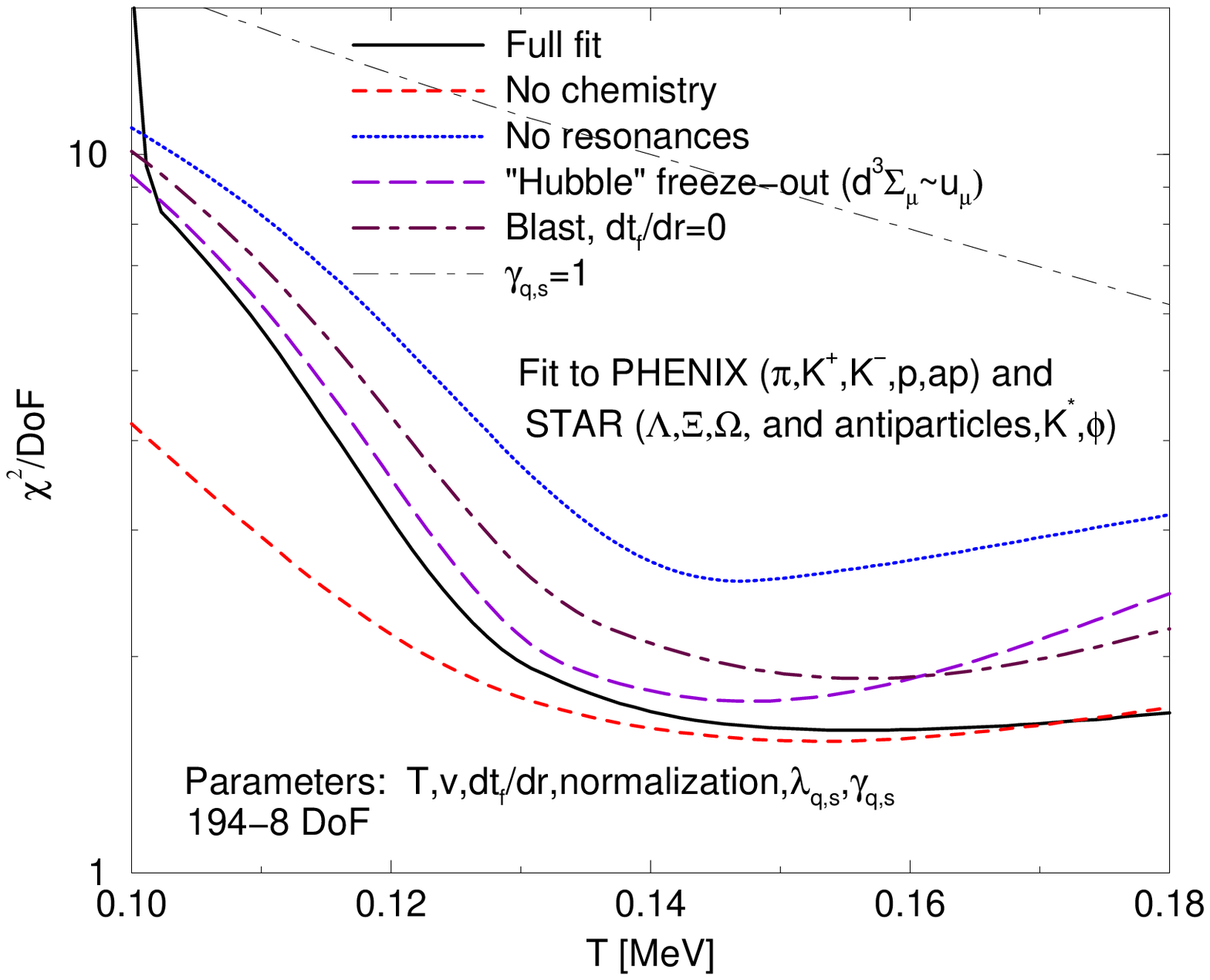}
}
\resizebox*{!}{0.247\textheight}{
\includegraphics{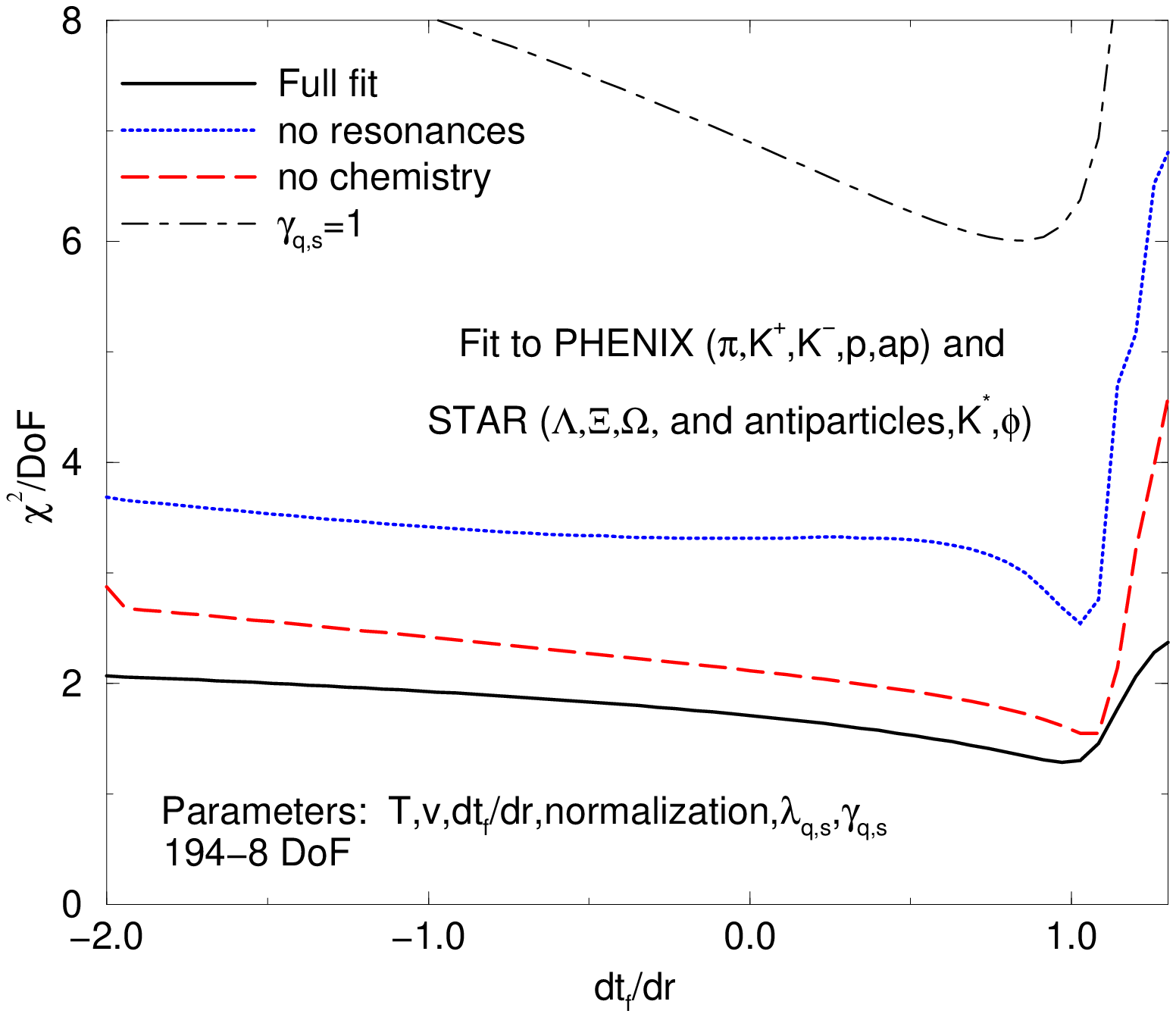}
}}
\end{center}
\caption{Fit to RHIC data\\
profiles for temperature (left) and fitted $\partial t_f/\partial r$ (right).   The models used in the fits are described in sections 1.1 and 1.2\\
Full fit includes fitted chemical potentials and $\gamma_{q,s}$, resonances and fitted $\partial t_f/\partial r$.
\label{fitdata} }
\end{figure*}
Finally, we have perfomed a fit to the available RHIC data.
The data sample we used is the same as the one used for the Montecarlo, but,
since the STAR and PHENIX spectra had different trigger requirements (notably
centrality) and acceptance regions, we have used two different
system volumes, one for STAR particles and another one for PHENIX.

The results are similar to the Monte Carlo data in many ways.
The $\chi^2/DoF$ was only slightly larger.\\
A fit  with particle-specific normalization gives a very similar $\chi^2$ and
fitted temperature to a fit including chemistry and resonances.
If chemistry is included in particle spectra analysis, resonances and
non-equilibrium are essential.
The fit to freeze-out geometry weakly points to
$\partial t_f/\partial r=1$, a picture that is supported by the
temperature $\chi^2$ profile, virtually identical to data sample III.
However, we can not claim our study to be complete in this respect, since we have not yet
investigated the effect of including
flow profile in the models.
As Montecarlo simulations have shown, the conclusion can differ
once these are taken into account.

\begin{figure*}[tb]
\begin{center}
\centerline{\resizebox*{!}{0.247\textheight}{
\includegraphics{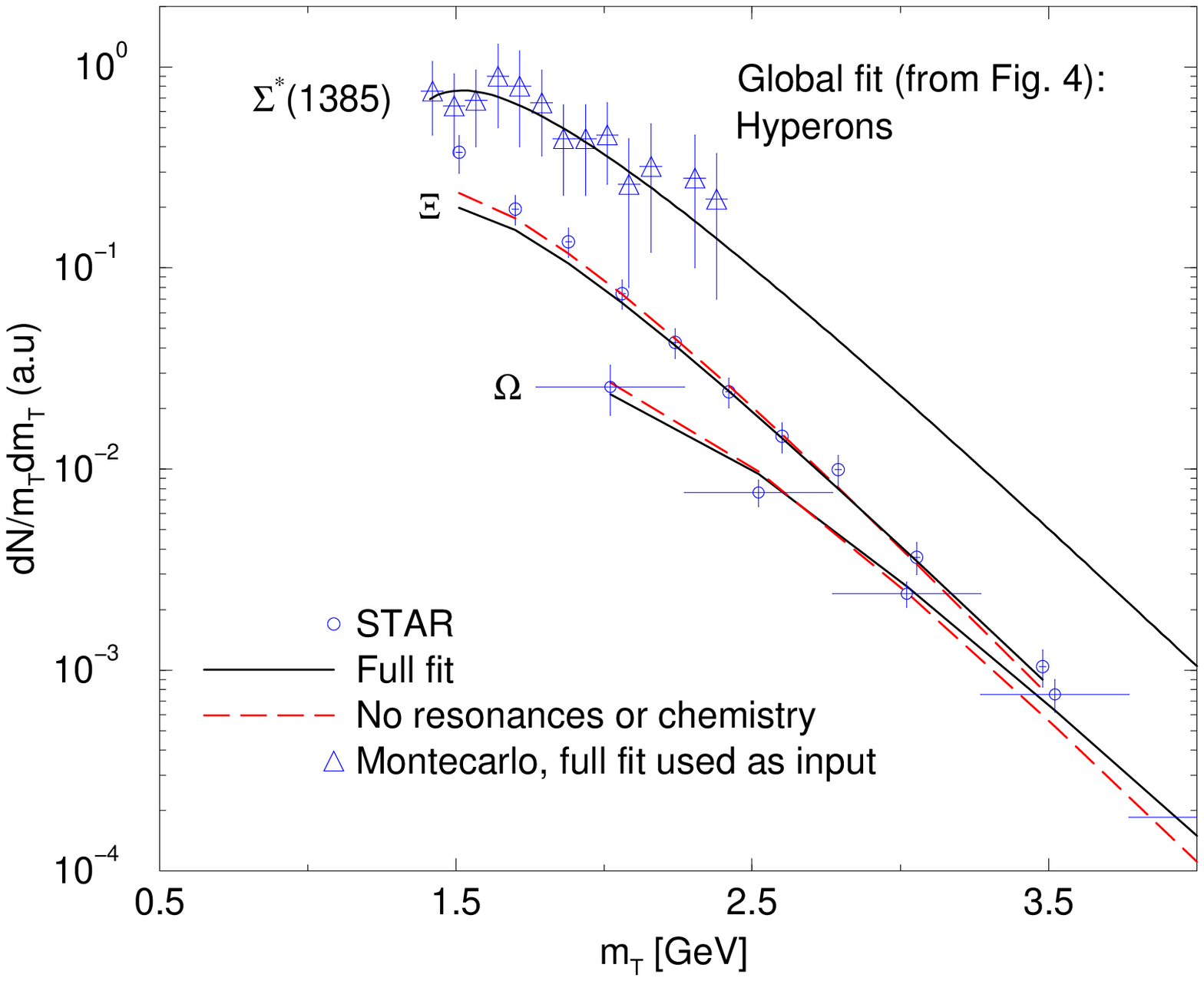}
}
\resizebox*{!}{0.247\textheight}{
\includegraphics{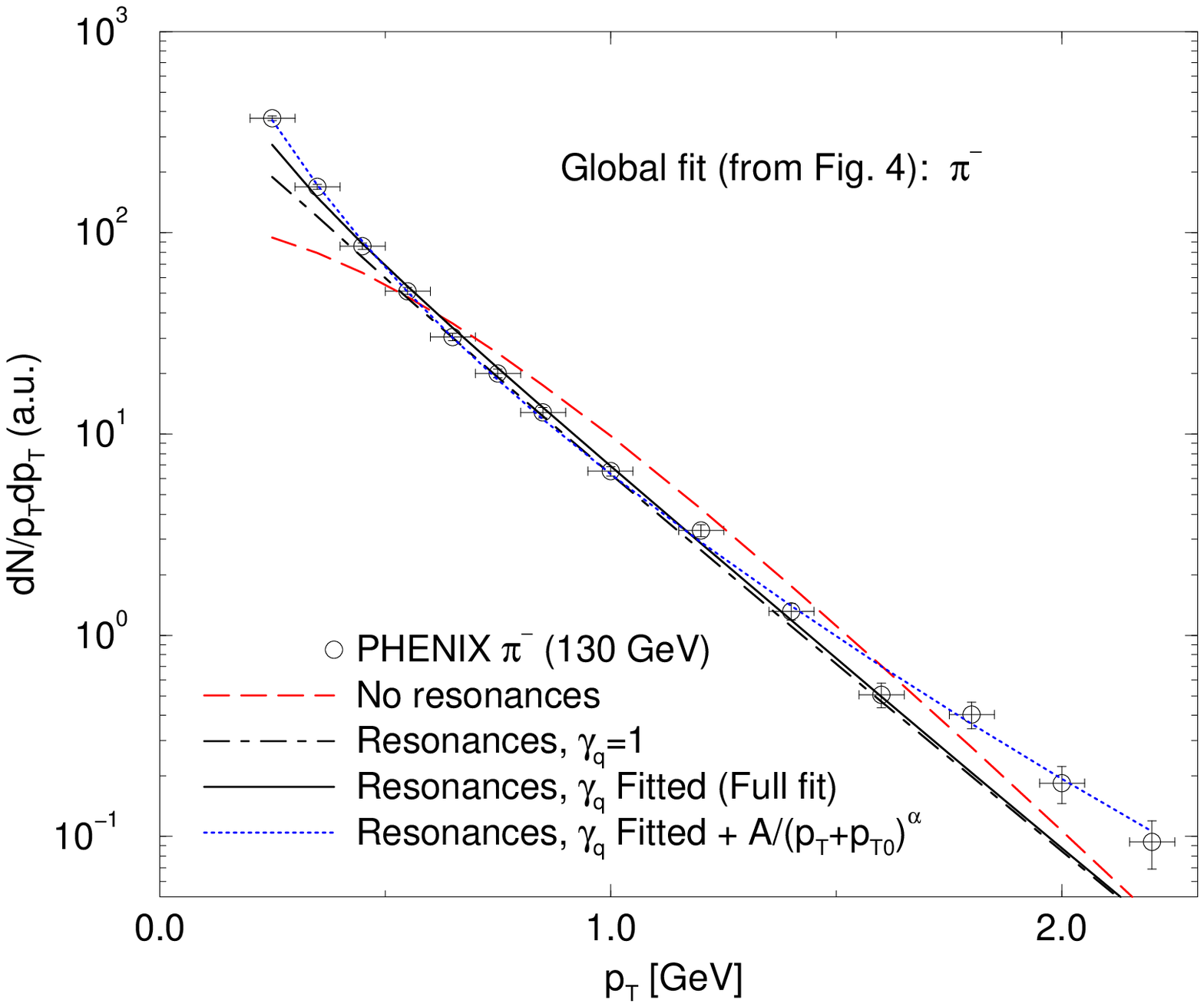}
}}
\end{center}
\caption{Left:  $\Xi$ and $\Omega$ $m_T$ distributions, together with the global fit (from Fig. \ref{fitdata}, solid black and dashed red).  The model with chemical potentials (solid black) was used to predict the
$\Sigma^*$ $m_T$ distribution using the Montecarlo.\\
Right: PHENIX $\pi^-$ $p_T$ distribution, with the global fit from Fig \ref{fitdata}.
As can be seen, both resonances and $\gamma_q$ help, but are not  sufficient to explain the pion distribution fully.
\label{twoplots} }
\end{figure*}

Fig. \ref{twoplots} shows the hyperon and pion spectra from the global fit 
of Fig. \ref{fitdata}.   A comparison of the fits on the left panel confirms that a model
with no chemistry is about as good at fitting particle spectra as a model
with resonances and chemical potentials.    However, the second fit also has
predictive power:  We have used the fitted parameters to predict the $m_T$ spectrum for the $\Sigma^*$.
Unsurprisingly, we found that the $\Sigma^*$ should have roughly the same slope as the $\Xi$s, but its total multiplicity should be about three times as big.
We therefore suggest that a greater sample of spectra, in particular more spectra of heavy resonances taken within the same centrality bin as light particles (to make sure
both flow and emission volume are the same for each particle)  would help in
establishing whether chemical potentials are a good way to normalize
hadron spectra or not.

The only spectrum which presents a significant systematic deviation for most
models is the $\pi^-$.    As Fig. \ref{twoplots} (right panel) shows, most models fail to catch the upward dip of the low momentum pions, and indeed simple transverse expansion predicts the reverse trend \cite{burward-hoy,van-leeuwen,NA57,castillo}.   Including resonances, and allowing
for $\gamma_q > 1$ helps (the latter is equivalent to postulating a pion
``chemical potential''  \cite{heinzpions})
However, to fully account for the lowest momentum pions, even addition of resonances and $\gamma_q > 1$ are not enough.
One has to add a power-law component to the pion spectrum
\begin{equation}
\label{powerlow}
E\frac{dN}{d^3 p}=\left(E\frac{dN}{d^3 p} \right)_{cf}+\frac{A}{(p_T+p_{T0})^{\alpha}}
\end{equation}
This contribution (roughly $6 \%$ of the total pion yield in the best fit) also accounts for the highest $p_T$ pions.
Such a parametrization has been justified \cite{powerlaw1,powerlaw2} and successfully used \cite{peitzmann} before.

In conclusion, we have presented a comparison of different statistical
models currently used to fit spectra in heavy ion collisions.
We have described how these different models arise from different freeze-out scenarios of a system hadronizing
from a thermalized quark gluon plasma.
We have used MonteCarlo simulated data to study the sensitivity to model choice of presently
available experimental data, and have evaluated different models ability to
fit presently existing RHIC data.
While data slightly favors a chemical non-equilibrium explosive freeze-out, there is not enough evidence to make a definitive conclusion
about this issue.   We hope the forthcoming 200 GeV results will clarify this further.
\ack
Supported  by a grant from the U.S. Department of
Energy,  DE-FG03-95ER40937\,.\\
GT acknowledges partial conference support
provided by NSF grant PHY-03-11859\\
We thank Magno Machado (IF-UFRGS) for helpful revision suggestions.

\end{document}